\documentclass[aps,prl,twocolumn,superscriptaddress]{revtex4}
\usepackage{graphics}
\usepackage{epsfig}

\begin{document}

\title{Functional alignment of regulatory networks:
A study of temperate phages}

\author{Ala Trusina}
\affiliation{Department of Theoretical Physics, Ume{\aa} University,
901 87 Ume{\aa}, Sweden}
\affiliation{NORDITA, Blegdamsvej 17, Dk 2100,
Copenhagen, Denmark}\homepage{cmol.nbi.dk}
\author{Kim Sneppen}\email{sneppen@nbi.dk}
\affiliation{NBI, Blegdamsvej 17, Dk 2100, Copenhagen, Denmark}
\author{Ian B. Dodd}
\author{Keith E. Shearwin}
\author{J. Barry Egan}
\affiliation{Discipline of Biochemistry, School of Molecular and Biomedical Science,
University of Adelaide, South Australia 5005, Australia}

\date{\today}

\begin{abstract}
The relationship between the design and functionality of molecular
networks is now a key issue in biology. Comparison of regulatory
networks performing similar tasks can give insights into how
network architecture  is constrained by the functions it directs.
We here discuss methods of network comparison
based on network architecture and signaling logic.
Introducing local and global signaling scores for
the difference between two networks
we quantify similarities between
evolutionary  closely and distantly related
bacteriophages.
Despite the large evolutionary separation
between phage $\lambda$ and 186
their networks are found to be similar when difference is
measured in terms of global signaling.
We finally discuss how network alignment can be used to
to pinpoint protein similarities viewed from the network perspective.
\end{abstract}

\maketitle
\section{Synopsis}
Networks of interacting genes and proteins orchestrate the complex
functions of every living cell.
Decoding the logic of these biochemical circuits is a central
challenge facing biology today.  Trusina et al.
describe a mathematical method for aligning two regulatory
networks based on their signaling properties, and apply it
to a case study of three bacteriophages, simple biological ``computers"
whose genetics are exceptionally well characterized.
The comparison reveals a surprising similarity between regulatory networks
of the creatures, even when they have very distant evolutionary relationships.
The method introduced here should be applicable to other networks,
and thus help illuminate the computational substructures of living
systems.

\section{Introduction}
The functioning of living organisms is based on an intricate network of
genes and proteins regulating each other. Various organisms differ due
not only to differences in the constituting components (genes/proteins),
but also because of the organization of these regulatory networks. It is
therefore important to address similarities and differences not only in
protein sequences but also in the interaction patterns of the proteins.
Thus, large scale analysis of protein-protein and protein-DNA
interactions have provided insight into the local design features of
subcellular signaling 
\cite{Jeong,Shen,Maslov2002-2};
network alignment based on sequence similarities permits
alignment of related motifs \cite{pathblast,Berg}.

Here we suggest to compare networks through an alignment method
that is based solely on network architecture and signaling logic,
and thus does not rely on sequence similarity of the involved proteins.

As a study case we consider the
regulatory networks of two very 
well-characterized temperate
bacteriophages of {\sl E. coli}, $\lambda$ and 186 (Fig. 1).
These two phages represent two distinct classes of temperate
bacteriophages: the lambdoid phages - which include $\lambda$, P22, 434, HK97
and HK022, and the P2 group - which includes P2, 186, HP1, K139 and
PSP3. $\lambda$ and 186, are not detectably related in sequence and have
different genome organizations. Using tBLASTx  \cite{blast} to compare all of the
reading frames, there are only two clearly homologous protein pairs -
the $\lambda$ endolysin R (P03706)/186 (PO80309) (E-score~=~$10^{-34}$) and a pair of
early lytic proteins of unknown function (E-score~=~2$\times 10^{-4}$). No
significant similarity was detectable at the nucleotide level (using
BLASTn, \cite{blast}). On the genome level, the arrangement of genes, promoters
and operators is very different \cite{Dodd2005,Portelli,Shearwin1998,Dodd2002}.
As a control of methodology, we also
consider the P22 phage, which as a member of the lambdoid family allows
us to compare topologies of evolutionarily related networks.

\begin{figure}
\centerline{\psfig{file=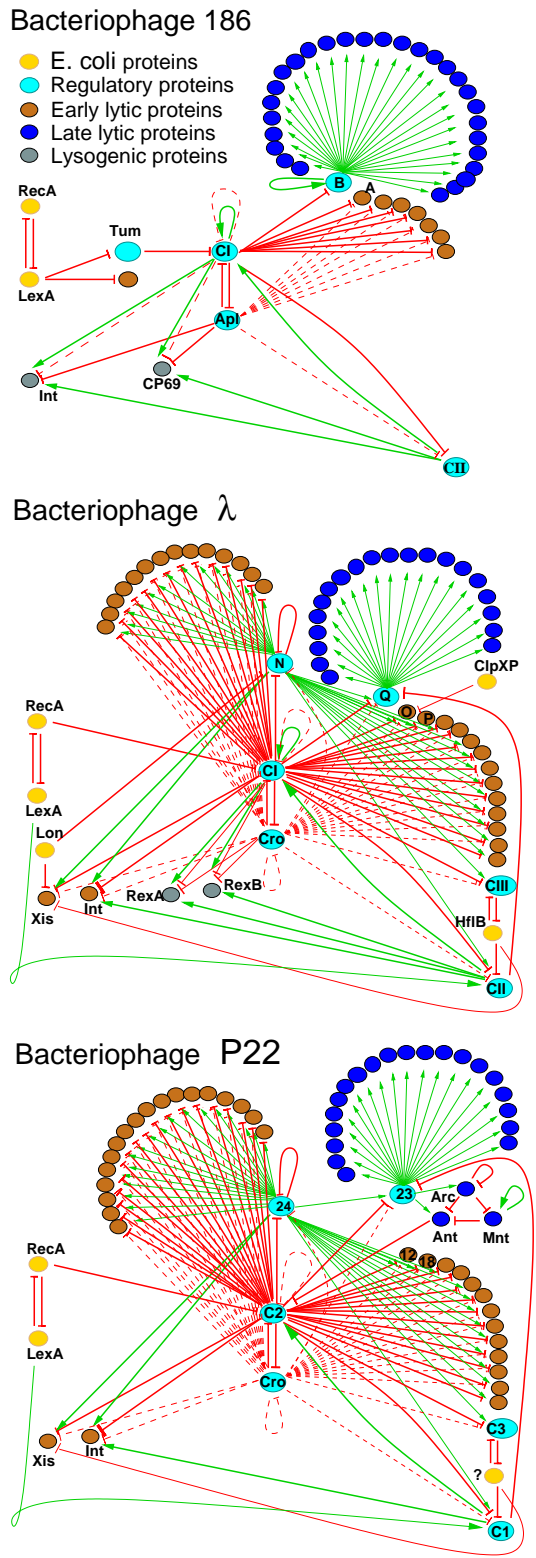,angle=0,width=7.0cm}}
\caption[]
{
The genetic regulatory networks for phage 186 for phage $\lambda$
and P22 all of which are temperate and infect
{\sl Escherichia coli}. The proteins are colored according to
their functions and expression mode in the lysis-lysogeny life
cycle of the phages. We summarize the influence of one protein on
another by either a green (positive, e.g., transcriptional
activation) or a red (negative, e.g., repression) arrow. The
dashed lines show relatively weak regulations. Database entry for
$\lambda$ genome is J02459, for 186 genome -- U32222, and for P22
genome it is NC$\_$002371. } \label{sneppen-fig1}
\end{figure}

As temperate phage, both 186 and lambda can be in two states: a lytic
state where many proteins are active in the replication of the phage DNA
and the construction and release of virus particles; and a lysogenic
state where the phage genome is integrated into the bacterial chromosome
and only a few proteins are active. For both phages, three core proteins
(CI (P03034), Cro (P03040) and CII (P03042) in $\lambda$, and CI (P08707), Apl
(P21681) and CII (P21678) in 186) do the main computations, with the
switch into lysogeny being coordinated by CII and the reverse switch
into the lytic mode initiated by activation of the host SOS response
protein RecA (P03017). The gene regulatory networks of all temperate
phage have evolved to provide lysogenic and lytic states, and more than
that, to switch from one state to another when particular signals have
been received from bacterial proteins, and thus effectively perform the
same function.

Given that 186
and $\lambda$ are both temperate, i.e. performing similar function,
but are evolutionary separated, we asked whether
we can detect structural similarities and what is the scale at which
these similarities are detectable~?

\section{Results}

\noindent
Visual comparison of the 186 and $\lambda$ networks (Fig. 1), suggests
both strong similarities but also major differences. One
way to quantify the similarity of two networks is by edit
distance \cite{Bunke}.  Assume that we know which nodes (here, proteins)
in network A and B should be paired.
For networks of the same size, we define edit distance
as the number of insertions or removals of edges (regulatory connections)
one has to perform on network A to obtain B.
This is quantified through
\begin{equation}
D_E(A,B) = \sum_{i,j}|A_{ij} - B_{ij}| \;,
\end{equation}
The elements $A_{ij}$ and $B_{ij}$ specify whether the direct regulation of
$i$ on protein $j$ is positive, negative or absent and are constructed such that
each element can keep both positive and negative links (for details
see eq. (2) below).

In case we do not know which nodes in networks A and B should be paired,
we find the optimal identification by minimizing $D_E$ as described in
{\sl Materials and Methods} section. 
This yields the minimal distance between the
networks, as well as an optimal alignment of the individual nodes.
This distance we call the edit difference.

The minimal edit difference between {\it related} phages is small
$D_E(\lambda,P22)=18$, compared to the larger scores for 
evolutionary separated phages, see table \ref{table1}. The
$D_E=18$ means that, the $\lambda$ network of $62$ proteins and
$144$ connections can be constructed by making $18$ edits of the
connections in a $62$ protein subset of the 67 protein $P22$
network (adding/removing a link is a single edit, changing the
sign of a connections needs two edits). To get an idea of the
significance of the obtained $D_E$ values, we compare with optimal
alignments of $500$ randomized versions of the two networks. The
randomization procedure was designed to conserve the local
properties of the networks in order to try to keep their general
biological features. Firstly, the core-hub topology common in
biological networks \cite{Maslov2002-2} was maintained by
conserving for each protein the number of its regulators (inputs)
and the number of proteins regulated by it (outputs). Secondly,
the number of each sign (positive and negative) of the input and
output connections was kept for each node.

\begin{figure}
\centerline{\psfig{file=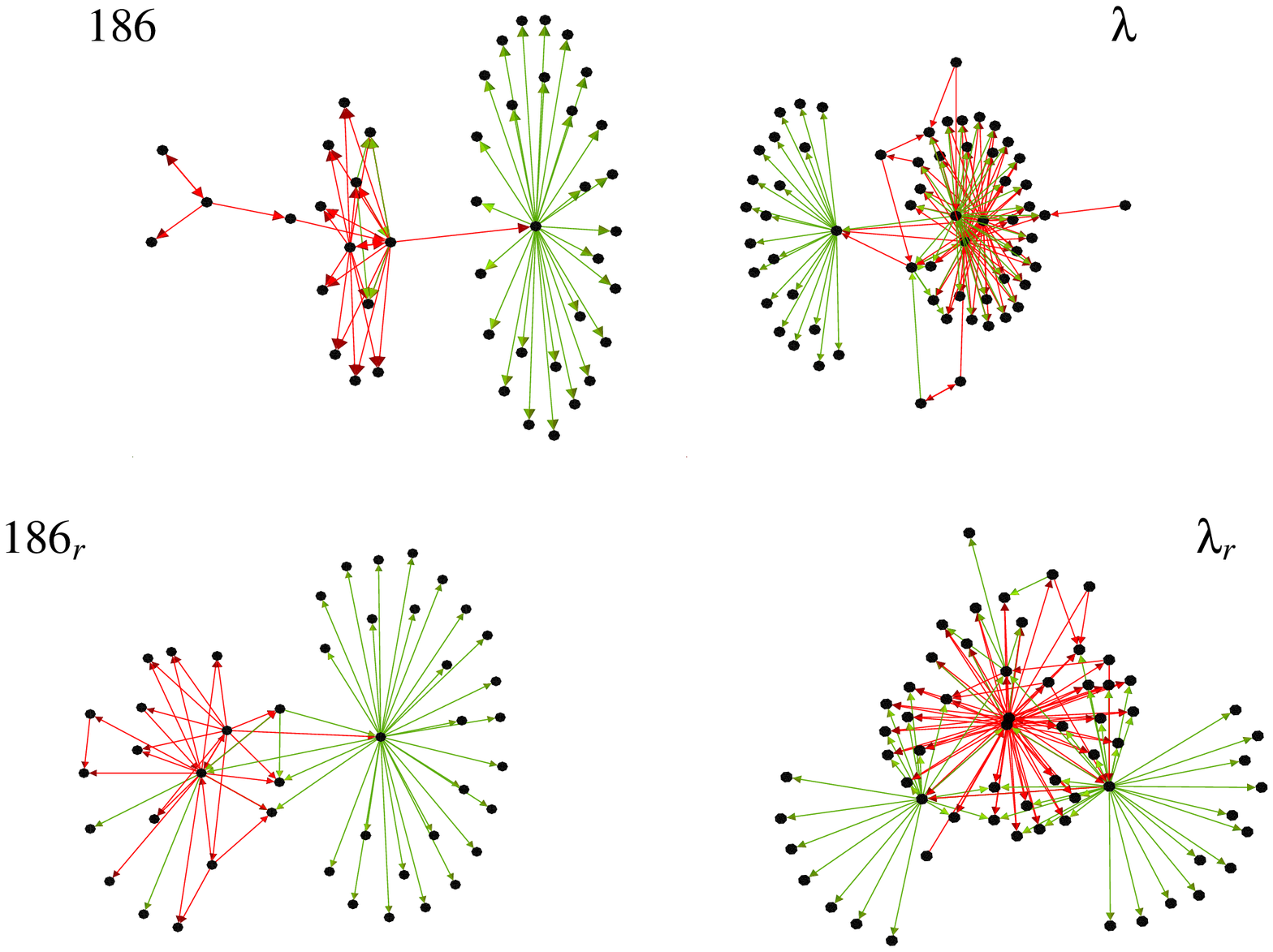,angle=0,width=8.0cm}}
\caption[]
{Illustration of the differences
between the real 186 and $\lambda$ networks (top) and an example of their
randomized counterparts (bottom). These examples of randomized networks
show that it possible to preserve local properties, yet obtain different
network structures.
} \label{sneppen-fig2}
\end{figure}

 The constrain of preserving the local properties does not fix
the network completely: while keeping the number of positively/negatively
regulated proteins one can still 
change which exactly of them are being regulated.
The structure of the resulting random
networks is rather different, as seen in the examples shown in Figure \ref{sneppen-fig2}.

Overall we find that $D_E$ scores between any pair of randomized
networks are similar. When comparing scores between real network,
with that of their random counterparts in table \ref{table1} one
see no clear trend. In particular the differences between these
randomized versions, $\lambda_r$ and $186_r$ were
indistinguishable from that of the real networks:
$D_E(186_r,\lambda_r) = 32 \pm 2$.

We reasoned that the functional similarity of networks
might be better reflected in a less local measure of functionality.
We therefore introduce a signaling difference $D_S$, which aims at capturing
both direct (as in $D_E$) and also indirect regulation through a sequence
of intermediate proteins. For each pair of
proteins ($i,j$), we consider whether $i$ sends a signal to $j$, and
if so whether the signal along the shortest path is
positive or negative. In this
spirit we define the sign of a signal as the product of the signs of
all links on the shortest path from $i$ to $j$.
An example where this procedure nicely reflects the functionality in terms of
its ``Boolean'' logic \cite{Kauffman} is found in the
pathway from RecA to CI in the two phages .
In $\lambda$, active RecA directly catalyzes
self cleavage of CI \cite{Little1984};
whereas in 186, RecA acts through the degradation of LexA (P03033),
that in turn represses the protein Tum \cite{Shearwin1998} (P41063),
which in the absence of repression binds CI and
prevents it from performing its function. Thus the simple -1 signal
in $\lambda$ is in 186 replaced by a signaling consisting of $(-1)
\times (-1) \times (-1)=-1$.
In other words repressing a repressor is effectively an activation.

Because the regulation of one protein by another may be positive through
one series of links and negative through another, two matrices were used
for each network, one for positive signals ($A^{S+}$ and $B^{S+}$) and one for
negative signals ($A^{S-}$ and $B^{S-}$). If the effect of protein $i$ on
protein $j$ is only positive, then $1$ is placed into
$A^{S+}_{ij}$ and 0 into $A^{S-}_{ij}$. If the effect is only negative,
then $0$ is placed into $A^{S+}_{ij}$ and 1 into $A^{S-}_{ij}$.
If there are positive and negative signals along paths of equal
length (e.g. from RecA to $\lambda$ CII via LexA or CI),
then 1 is placed into both matrices.
Observe that when positive and negative signals come
to the same node, they are not canceling each other.
This is intended, as often signals will arrive
at different times or at different conditions \cite{footnote}.

The signaling difference between two
networks A and B is then defined as
\begin{equation}
D_S (A,B) \; = \; \sum_{ij} \; | A^{S+}_{ij} - B^{S+}_{ij} | \;+\; | A^{S-}_{ij} - B^{S-}_{ij} |
\end{equation}
which takes into account differences in both positive
and negative signaling along the shortest paths between any pair of nodes.
Like $D_E$, the minimum difference $D_S$ is calculated by optimizing which proteins in 186 should be
identified with which proteins in $\lambda$,
and in addition which $\lambda$ proteins should be excluded.
Excluding a protein means that the signaling to and from that protein is not counted in $D_S$,
whereas signaling across the excluded protein is included.

\begin{table}[t]
\begin{tabular}{||c||c|c||c|c||}
\hline
netA, netB & $D_E$ & $P_E$ & $ D_S$ & $P_S$ \\
\hline
$\lambda$, 186  & 33 & & 43& \\
$\lambda_r$, $186_r$  & 32 $\pm$ 2 & 0.27 & 109 $\pm$ 33&0.01 \\
\hline
$\lambda$, $P22$  & 18 & & 106& \\
$\lambda_r$,$P22_r$  & 33 $\pm$ 4 & 0.00 & 255 $\pm$ 55& 0.00 \\
\hline
$P22$, 186  & 25 & & 97 & \\
$P22_r$, $186_r$  & 31 $\pm$ 1 & 0.00 & 161 $\pm$ 36 &0.03\\
\hline
\end{tabular}
\caption{The overall difference measures
$D_E$, $D_S$ between the networks,
with respective P-scores as defined in text.}
\label{table1}
\end{table}

Optimizing protein alignment based on signaling, we find
$D_S(186,\lambda )=43$.
Again, the significance of this difference was determined by repeatedly performing randomization of the networks as
described above, creating the $A^{S+}$ and $A^{S-}$ matrices and obtaining the
minimal $D_S$. The differences between random networks,
$D_S(186_r,\lambda_r)=109\pm 33$, is much larger than between the real networks.
This is further quantified by a P-score, $P(D_S>D_S(random))=0.01$,
defined as the probability that two randomized networks
will have a smaller difference than that between the real networks.

Thus all three networks are similar in their signaling pattern.
To confirm that this signaling similarity is not
generally conserved among biological networks, we have
compared the phage networks with other networks that
perform different functions (e.g. the {\sl S. cerevisiae}, \cite{Li},
cell cycle network and the {\sl B. subtilis} competence
network, \cite{Hamoen}). We found that $D_S$ is much
larger and the P-scores are close to 1 in these alignments, indicating
that the low signaling difference between the phage networks is a special
property of these functionally similar networks.\\

We have also considered other variants of the difference measures,
in particular including all non-repetitive paths between pairs
of proteins, with all paths weighted equally.
In that case we also find that $D_{S-all}(\lambda,186)=390$
between real network is smaller than
$D_{S-all}(\lambda_r,186_r)~=~583\pm122$ between the randomized counterparts.
Also, using the shortest paths, we have investigated
differences  between networks where weak links (the dashed ones in Fig. 1)
are weighted less (by a factor 0.5 or removed altogether).
$D_S$ scores between networks got smaller,
but overall significance remained similar.

\section{Discussion}

The pathway related $D_S$ score allowed us to identify significant
similarity between two very distantly related biological
networks , see table \ref{table1}. In contrast, the edit
difference measure, which looks only at the local wiring
structure, is sometimes blind to this more global ``homology''.
Thus although edit difference partially captures network
similarities through a patchwork of local matchings, it is less
sensitive to pathway disruptions.

It is not clear whether the functional similarity between the
lambda and 186 networks detected by the $D_S$ measure is a result
of convergent evolution or is a remnant of a shared ancestral
network. Under either scenario it is clear that the two network
structures must be strongly constrained by functional
requirements, given the evolutionary separation of the two phages.
A potential bias should be noted here: knowledge of the three
phage networks is not complete, even for $\lambda$, and it is thus
possible that some of the observed similarity in the networks is
due to knowledge of connections in one phage network having
influenced the discovery of connections in the others.

\begin{figure}
\centerline{\psfig{file=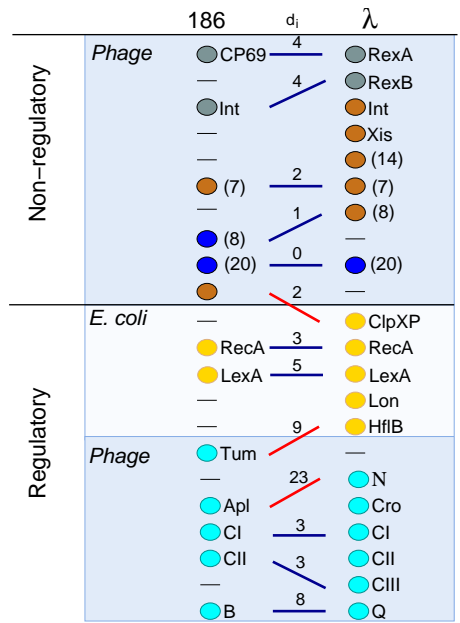,angle=0,width=6.0cm}}
\caption[]
{
Alignment of two phage networks. Placement of proteins is based on
our knowledge
\protect{\cite{Shearwin1998,Dodd2005,Portelli,Little1984,Kobiler}}
and the lines connecting them are associated to the minimal $D_S$
alignment. Proteins that perform similar functions or are
regulated similarly are placed on the same level; thus horizontal
lines mark ideal matching. Blue lines correspond to meaningful
alignments, red lines are the misalignments.  The numbers above
the lines, $d_i$, reflect the differences in signaling between the
aligned proteins and are the contributions to the minimal
difference $D_S=\frac{1}{2}\sum_i{d_i}=43$. The numbers in the
parentheses indicate multiple equivalent proteins, making the sum
of all shown signaling differences equal to $2\cdot 43$. The key
regulators RecA, LexA and CI are identified correctly whereas the
misidentification of CII with CIII is reasonable since both favor
entry into lysogeny through the same pathway. The major
discrepancy is associated to different roles of Cro and Apl during
lysis (the weak links from Cro to Q and N in $\lambda$). }
\label{sneppen-fig3}
\end{figure}

The $D_S$ alignment allows us to address the role of various
proteins in pathway disruptions.
Figure \ref{sneppen-fig3} line up the $\lambda$ and 186 proteins
on the basis of pre-existing knowledge of their function or
mode of expression and have indicated the optimal $D_S$
alignment and the contribution of each pair to
the signaling difference.
The two alignments show good matches for late lytic genes as well
as for the regulators CI, CII and B from 186 aligned with CI, CII
and Q in $\lambda$. Thus in general 
functions of proteins in one
network teaches us about protein properties in the other network.
The lack of a good match between Apl (in 186) and Cro (in
$\lambda$), is due to the weak links from Cro, and reflects a
different functional role of Cro and Apl in the late lytic
development of phages. Insisting on alignment of Cro with Apl
results in $D_S=219$,  thus emphasizing the particular role of Cro
as a repressor of late lysis in $\lambda$.

Comparison of molecular networks is becoming an important element
of modern systems biology, both with regards to predictions of
eventual missing links \cite{Albert}, and for increasing our
understanding of functionality of information processing in the
networks. The here presented alignment methods address the
similarities on a local, respectively larger scale, associated to
signaling across networks.

In this regards we found that evolutionary
relationships ($\lambda-P22$) imply similar local regulation, with
low $D_E$ score.
For all temperate phages,  evolved to do similar
"computation",
 their regulatory networks
are found to be similar when viewed from a more global perspective
where both direct and indirect signals are included (low $D_S$ score
compared to random expectation).
 Thus the mechanistic and structural differences
on the scale of genome and promoter organization
disappear when considering the large scale
of the protein regulatory networks.
Going beyond immediate regulations allows
to capture functional similarity in the most robust way.

\section{Materials and Methods}
The present papers is based on the data on three bacteriophages
$\lambda$ (accession no. J02459), P22 (NC$\_$002371) and 186(U32222).
The regulatory networks were compiled from these database entries and
various literature sources: $\lambda$
(\cite{Hendrix1983,Dodd2005,Kobiler2004,Kobiler} and references therein),
for 186 (\cite{Portelli,Shearwin1998,Dodd2002} and references therein),
for P22 (\cite{Pedulla} and references therein).

In the {\sl Results} section we define two differences scores, $D_E$ and $D_S$
between a pair of networks $A$ and $B$.
Provided that we know which proteins in $A$ should be identified with which in $B$,
the scores are calculated as in Eq.~1 respectively Eq.~2. 
In case we do not know which nodes in networks $A$ and $B$ should be paired,
we need to find the optimal identification of nodes between them.
To do so, we define an alignment procedure
through the Metropolis algorithm \cite{Metropolis} designed to
reach the minimal distance $D$ between the networks:
Given two nodes and their corresponding partners in the other network
the elementary step is to switch partners and reevaluate the distance.
Iterating this procedure and using
simulated annealing \cite{Kirkpatrick} the method converges to
a global minimum.

If the two networks are of different size we
count only the contribution from a number of nodes given by
the smaller of the two networks.
In the larger network these nodes are selected to minimize
the distance using the above algorithm.

We would like to note that the above method is not intended
to reflect any evolutionary process, but is used to find the
optimal mapping of pairs of proteins that look similar from the
network perspective. The method is limited by the network size,
and in practice works for networks below 200 nodes.

The realization of the alignment algorithm in form of the Java applet
is available at http://www.cmol.nbi.dk/models/compar/compar.html.   

\section{Acknowledgment}
We warmly thank S. Brown, S. Krishna and S. Strogatz for constructive comments on the manuscript.
The work was supported by the Swed. Res. Council Grants
No. 621 2003 6290, 629 2002 6258 and ``Danmarks Grundforskningsfond"
through the center ``Models of Life" at the Niels Bohr Institute.\\
Work in the Egan lab is supported by the US NIH.


\begin{thebibliography}{10}
\bibitem{footnote}
An example of this is the two paths from RecA to
CII over respectively CI and LexA,
of which only the RecA-LexA-CII path is activated during lysis.
\end{thebibliography}
\end{document}